\documentclass[%
 reprint,
amsmath,amssymb,
prl,
superscriptaddress,
]{revtex4-2}
\usepackage{comment}
\usepackage[utf8]{inputenc}
\usepackage[T1]{fontenc}
\usepackage{mathptmx}
\usepackage{etoolbox}
\usepackage{lipsum}%
\usepackage{multirow}
\usepackage{graphicx}
\usepackage{dcolumn}
\usepackage{bm}

\begin{document}

\preprint{APS/123-QED}

\title{Incompressible Squeeze-Film Levitation}

\author{Mostafa A. Atalla}
\email{Corresponding author.
\\
m.a.a.atalla@tudelft.nl
\\
}
 \affiliation{BioMechanical Engineering Department, Delft University of Technology, The Netherlands}
\affiliation{
Cognitive Robotics Department, Delft University of Technology, The Netherlands
}
\author{Ron A.J. van Ostayen}
 \affiliation{Precision and Microsystems Engineering Department, Delft University of Technology, The Netherlands}
 
\author{Aimée Sakes}
 \affiliation{BioMechanical Engineering Department, Delft University of Technology, The Netherlands}

\author{Michaël Wiertlewski}
\affiliation{
Cognitive Robotics Department, Delft University of Technology, The Netherlands
}

\date{\today}

\begin{abstract}
Transverse vibrations can induce the non-linear compression of a thin film of air to levitate objects, via the squeeze film effect. This phenomenon is well captured by the Reynolds' lubrication theory, however, the same theory fails to describe this levitation when the fluid is  incompressible. In this case, the computation predicts no steady-state levitation, contradicting the documented experimental evidence. In this letter, we uncover the main source of the time-averaged pressure asymmetry in the incompressible fluid thin film, leading the levitation phenomenon to exist. Furthermore, we reveal the physical law governing the steady-state levitation height, which we confirm experimentally.
\end{abstract}

\maketitle

When an object is placed in closed proximity to a surface vibrating at sufficiently high frequency, it levitates. Levitation has been exploited in a wide range of applications from squeeze-film bearings \cite{Wiesendanger2001,Ide2005}, contactless manipulation and transportation of objects \cite{Ueha2000,Gabai2019} to friction-modulation in surface haptics touchscreens \cite{Winfield2007,Wiertlewski2016}. 

In all those examples, levitation occurs when the surrounding fluid is air, which is compressible, via the so-called squeeze-film effect. Having the same ability to levitate objects in incompressible liquid environments could help bring non-contact manipulation and friction modulation to a range of applications, including the medical field where blood and liquids are omnipresent. 

In-liquid levitation has been addressed by a limited number of studies. Hatanaka et al. \cite{Hatanaka1999} experimentally demonstrated that underwater squeeze-film levitation is possible. Nomura et al. \cite{Nomura2007} realized a noncontact transportation underwater using ultrasonic traveling waves. However, neither of these studies provided an explanation of the physical underpinnings of this phenomenon, with the second speculating that it might be due to the non-linear viscosity of liquid. Tamura et al. \cite{Tamura2006} hypothesized that underwater levitation is due to formation of cavitation bubbles, which collapse at the surface of the levitated object. Although their experimental data matches well with their hypothesis, the relatively large experimental film thickness ($>100~\mu m$) and high power transducer (350~W) limits the generality of their conclusion. This evidence indicates a gap in understanding the physical principles behind in-liquid levitation and thus, requires revisiting the existing theories.

The first leading theory that models squeeze-film levitation is the Reynolds' lubrication theory \cite{Reynolds1886}. This theory approaches the problem from the perspective of viscous fluids in a flow regime with  negligible fluid inertia. Langlois \cite{LANGLOIS1962} mathematically formalized the theory for the isothermal squeeze-film case, which was later adopted and validated experimentally by Salbu \cite{Salbu1964} who showed that air squeeze films levitate mainly due to the non-linear compressibility of the viscous air film. Clearly, such an explanation is not applicable anymore once a liquid is used. As a consequence, employing the Reynolds equation to model the in-liquid levitation phenomenon, fails and predicts no steady state levitation force. Based on this result, Stolarski et al. \cite{Stolarski2006} believed that oil squeeze films have no load-carrying capacity. However, this theoretical result clearly contradicts the experimental evidence, which disqualifies the lubrication theory in its current form from modeling liquid squeeze-film levitation.

The second leading theory in modeling squeeze film levitation is the acoustic radiation pressure theory. This theory takes the perspective of wave propagation in compressible inviscid fluids. The pioneering work of Chu and Apfel \cite{Chu1982} shed the light on the radiation pressure of compression waves 
acting on perfectly reflecting surfaces. This fundamental work was adopted and simplified later by Hashmimoto et al. \cite{Hashimoto1996} to model the special case of air thin films. Zhao et al. \cite{Zhao2013} showed experimentally, however, that this theory fails to capture the physics in air thin films of a typical thickness ($<100~\mu m$), which was further confirmed by the experimental results of Li et al. \cite{Li2022}. This is mainly because the boundary layer thickness is in the same order of magnitude of the film thickness, suggesting significant viscous effects \cite{Andrade2020}. The viscous effects of the boundary layer become even more significant in liquids whose viscosity is around two orders of magnitude higher, which rules out the applicability of this theory to model in-liquid squeeze film levitation.

A limited number of studies attempted to derive a unified viscoacoustic theory that works across the viscous and acoustic regimes. Melikhov et al. \cite{Melikhov2016} developed a viscoacoustic model and identified the different operating regimes for air squeeze films as a function of the levitation height, confirming a purely viscous regime for typical squeeze-film levitation systems. Ramanarayanan et al. \cite{ramanarayanan2022} proposed another unified theory which described critical parametric conditions that causes levitation forces to switch to adhesion forces in air squeeze-film systems. Remarkably, in the incompressible limit, their formulation predicted only weak adhesive squeeze-film forces, adding even more uncertainty around the behaviour of incompressible squeeze films.

In other terms, the existing theories fail to capture the physics of in-liquid squeeze-film levitation. We can also conclude that a viscous fluid approach (i.e. similar to the lubrication theory) is essential to tackle this problem given the comparable size of the film thickness and the boundary layer. However, relaxation of the assumptions of the lubrication theory is needed to uncover the underlying physics behind this phenomenon and find out its physical governing law.

In this letter, we show that a stable steady-state squeeze-film levitation of objects can be obtained in incompressible liquid environments. We uncover the fundamental pressure-inducing mechanisms in thin  films. Finally, We reveal the physical law that governs the steady-state levitation height, which we validate experimentally.

\begin{figure}
  \includegraphics{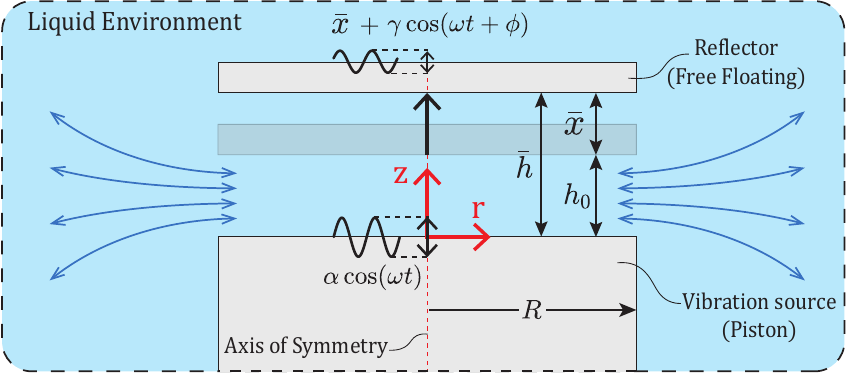}
  \caption{In-liquid squeeze film levitation. Initially, a film thickness $h_{0}$ separates the vibrator and the free-floating reflector. This initial film thickness originates from various sources such as the roughness of the two surfaces or misalignment. Upon the start of the vibration $\alpha \cos(\omega t)$, a pressure builds up in the film layer, which pushes reflector away distance $x$ until it reaches equilibrium at the time-averaged steady-state levitation distance $\bar{x}$. The reflector oscillates around its equilibrium position harmonically, denoted by $\gamma \cos(\omega t +\phi)$. The over-pressure in the liquid layer is linked to the steady state time-averaged levitation film height.}
  \label{problem_definition}
\end{figure}

Consider an axisymmetric system of a sinusoidal vibration source, a free-floating reflector and a thin film of liquid in between as shown in Fig.\ref{problem_definition}. We define the time-dependent film thickness $h(t)$ such that $h(0)=h_{0}$, where $h_{0}$ is an assumed initial film thickness between the source and reflector. The disc radius is $R$ such that $R\gg h(t)$. The source oscillates with an angular frequency $\omega$ and an amplitude $\alpha$ such that $\alpha < h(t)$. We define the squeeze Reynolds number \cite{Brunetiere2020} to be $Re_s = \rho\omega \bar{h}^{2}/\mu$ where $\bar{h}$ is the steady-state time-averaged film thickness, $\rho$ is the density of the liquid and $\mu$ is its viscosity. This number gives a measure of the relative significance of inertial and viscous effects. 

One of the key assumptions of the lubrication theory is that fluid inertia is negligible and thus fluid behaviour is dominated by viscous effects. This proved to be true for gas squeeze films since the squeeze Reynolds number is typically less than unity ($<1$) \cite{Minikes2004,Stolarski2008}. However, in the case of a typical liquid, such as water, the order of magnitude of the physical parameters of the Reynolds number is as follows: $\rho \approx O(10^3)$~Kg/m$^3$, $\omega \approx O(10^4)$~rad/s and $\mu \approx O(10^{-3})$~Pa.s. Given the experimental data of Hatanaka et al. \cite{Hatanaka1999}, we can also expect the order of magnitude of ($\bar{h}$) to be $\bar{h} \approx O(10^{-5})$~m. This yields a Reynolds number of order $Re_s \approx O(10^{1})$ suggesting the significance of the film inertia and thus, the lubrication theory's assumption becomes invalid. Therefore, we revisit the basic fluid governing equations, which take inertia into account.

Given the aforementioned axisymmetric system of a source and reflector and given that $R \gg h(t)$, we can safely assume that the pressure gradient across the film thickness is negligible compared to the radial pressure gradient ($\frac{\partial p}{\partial z} \ll \frac{\partial p}{\partial r}$) and therefore the pressure is only a function of the radial coordinate ($\frac{\partial p}{\partial z} = 0$) \cite{TICHYJA1970}.  With this assumption, the conservation of momentum and mass equations for this system using cylindrical coordinates, are given by \cite{TICHYJA1970, Hamrock2004}:
\begin{equation}\label{eq: navier-stokes}
    \rho\left(\frac{\partial v_{r}}{\partial t}+v_{r}\frac{\partial v_{r}}{\partial r} +v_{z}\frac{\partial v_{r}}{\partial z}\right) = -\frac{\partial p}{\partial r} + \mu \frac{\partial^{2} v_{r}}{\partial z^2}
\end{equation}
\begin{equation}\label{eq: continuity}
    \frac{1}{r} \frac{\partial}{\partial r}(rv_{r})+ \frac{\partial v_{z}}{\partial z} = 0 
\end{equation}
where $p$ is the liquid pressure, $\rho$ is its density, $\mu$ is its viscosity, and $v_{r}$ and $v_{z}$ are the liquid velocity fields along $r$ and $z$ respectively. To solve for the pressure and velocity fields, we follow the iterative scheme of Kuzma and Jackson \cite{Kuzma1967,JACKSON}; we first approximate the inertial forces (left hand side of Eq.\ref{eq: navier-stokes}) by employing the velocity profiles of the classical lubrication theory, which are given by:  
\begin{equation}\label{eq: velocity-profile-lubr1}
    v_{r} = \frac{3r\dot{h}}{h^{3}}\left(z^{2}-hz\right) \tag{3a}
\end{equation}
\begin{equation}\label{eq: velocity-profile-lubr2}
    v_{z} = -\frac{\dot{h}}{h^{3}}\left(2z^{3}-3hz^{3}\right) \tag{3b}
\end{equation}

Then, by integrating Eq.\ref{eq: navier-stokes} twice with respect to ($z$) and assuming static boundary conditions ($v_{r}(r,0,t)= v_{r}(r,h,t) = 0$), we find the following new expression for the radial velocity field ($v_{r}$):

\begin{multline}\label{eq: velocity profile new}
    v_{r} = \frac{1}{\mu}\frac{\partial p}{\partial r}\left(\frac{z^2}{2}-\frac{hz}{2}\right) + \frac{\rho}{\mu} \left[ \frac{3r\ddot{h}}{h^3}\left(\frac{z^4}{12}-\frac{hz^3}{6}+\frac{h^3z}{12}\right) \right. \\
   \left. +\frac{r\dot{h}^2}{h^6}\left(-\frac{z^6}{10} + \frac{3hz^5}{10} - \frac{3h^2z^4}{4} +h^3z^3  - \frac{9h^5z}{20}\right)\right] \tag{4}
\end{multline} 

The conservation of mass principle requires that the inflow or outflow across the control volume of the film is equal to the volume change due to the source vibration and the reflector levitation. This condition can be expressed mathematically in integral form as follows: 

\begin{equation}\label{eq: continuity integral}
    \int_{0}^{h} v_{r} dz= - \frac{r \dot{h}}{2} \tag{5}
\end{equation}

Finally, by substituting the velocity profile expression Eq.\ref{eq: velocity profile new} into the continuity equation Eq.\ref{eq: continuity integral}, we obtain the following expression for the pressure gradient along ($r$) \cite{Kuzma1967}:
\begin{equation}\label{pressure-gradient}
    \frac{\partial p}{\partial r} = \frac{6\mu r \dot{h}}{h^3} + \frac{3\rho r \ddot{h}}{5h} - \frac{15\rho r \dot{h}^2}{14h^2} \tag{6}
\end{equation}

This expression has different coefficients of the second and third terms compared to the work of Li et al. \cite{Li2010}, because in that work the term ($v_{z}\frac{\partial v_{r}}{\partial z}$) was neglected despite being comparable in magnitude to ($v_{r}\frac{\partial v_{r}}{\partial r}$) \cite{Kuzma1967}. 
This first order iterative solution was shown previously to be stable and to agree with the full numerical solutions \cite{Grimm1976}, unlike the higher order solutions \cite{TICHYJA1970}.
By integrating the pressure gradient expression with respect to ($r$) assuming the boundary condition ($p(R,t)=p_{a}-\Delta p$) where $p_{a}$ is the ambient pressure and $\Delta p$ is a pressure loss term due to the edge effect, we obtain the following pressure field expression:

\begin{equation}
\begin{aligned}\label{eq: pressure-field}
&p-p_{a}=\frac{r^2-R^2}{2} \left( \frac{6\mu \dot{h}}{h^3} + \frac{3\rho \ddot{h}}{5h} - \frac{15\rho \dot{h}^2}{14h^2} \right) - \Delta p
\\
&\text{where } \Delta p=
\begin{cases}
\frac{C_{e} \rho R^2 \dot{h}^2}{8 h^2}& \text{if } \dot{h} > 0\\
0              & \text{otherwise}
\end{cases}
\end{aligned}\tag{7}
\end{equation}

This pressure profile expression suggests three main pressure-inducing mechanisms for incompressible films. The first one is associated with the viscosity of the liquid represented by the term $6\mu \dot{h}/h^3$. This term is symmetric and thus, has no contribution to the time-averaged pressure. The second and third mechanisms are associated with the temporal and convective accelerations of the liquid squeezing in and out of the film, represented by the asymmetric terms $3\rho \ddot{h}/5h$ and $15\rho \dot{h}^2/14h^2$ respectively. In addition, accounting for liquid inertia results in an inevitable pressure drop $\Delta p$ at the edge of the squeeze film during negative squeeze motion ($\dot{h}>0$), due to the sudden contraction of the liquid at the film entrance. Using Bernoulli's equation, we can find an expression for the pressure drop term \cite{hashimotoH1995,Hori2006} equal to ${C_{e} \rho R^2 \dot{h}^2}/{8 h^2}$ where $C_e$ is a pressure loss coefficient that depends mainly on the geometry of the film entrance (i.e. sharp or round edge). This edge effect term along with the asymmetric temporal and convective acceleration terms contribute to a non-zero time-averaged squeeze-film pressure 
(refer to the supplemental material \cite{supp} for more details about the pressure terms).


To account for the levitated object being free floating, we assume a single degree of freedom system where the reflector is a mass. The mass is connected to the vibrating surface through a liquid layer whose initial thickness is $h_{0}$. We introduce two independent coordinates ($x, y$) where $x$ is the displacement of the reflector and $y$ is the displacement of the vibrating surface.  The film height $h(t)$ is a function of $x$ and $y$ coordinates as follows: 

\begin{equation}
    h=h_{o}+x-y \text{   ,   } \quad \dot{h}=\dot{x}-\dot{y} \text{   ,   } \quad  \ddot{h}=\ddot{x}-\ddot{y}
    \label{eq: film thickness}\tag{8}
\end{equation}

\begin{figure}
    \centering
    \includegraphics{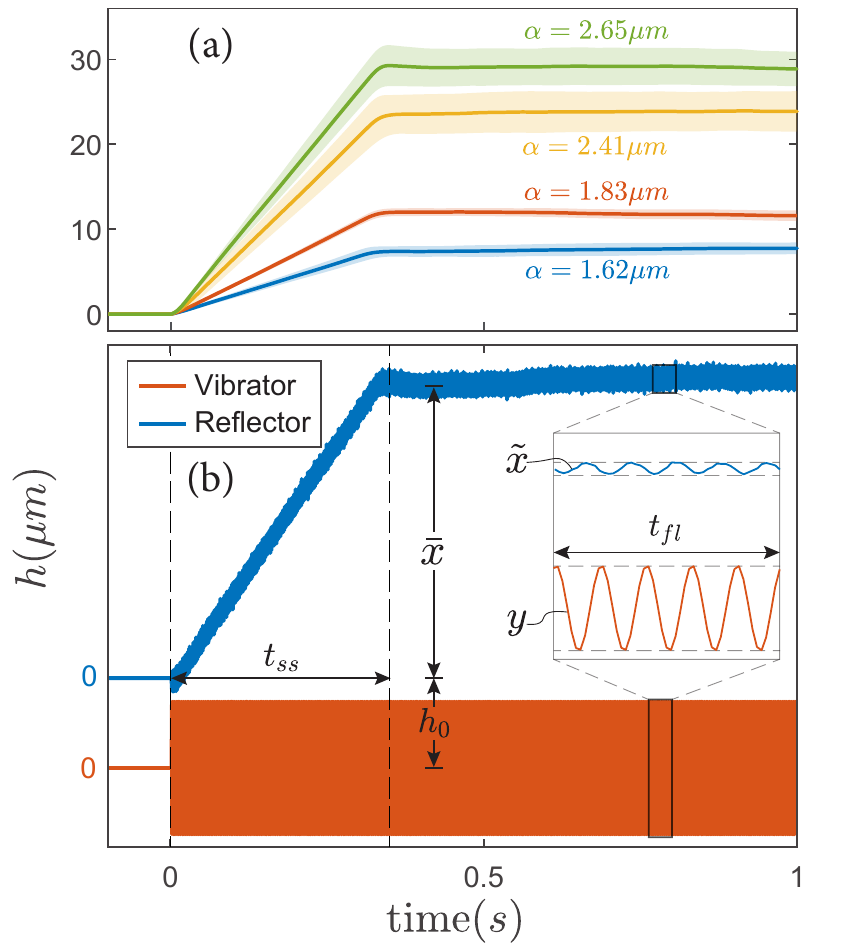}
    \caption{(a) Demonstration of the stable steady-state levitation of a free-floating reflector of mass $64g$, at $40kHz$ excitation. For each vibration amplitude, five trials of measurements were collected and the results are plotted as the mean displacement (solid lines) and the standard deviation (shades). (b) A detailed view of one experimental trial corresponding to the input vibration amplitude $\alpha=1.62 \mu m$, with the vibrator displacement being $y$. We observe that the floating mass levitates to a steady-state position $\bar{x}$ and oscillates around the equilibrium position, where the oscillations are represented as $\tilde{x}$. 
    The response shows two characteristic time scales: (a) time corresponding to the mass reaching steady-state levitation distance $t_{ss}$ (b) time associated with the reflector oscillations $t_{fl}$.}
    \label{fig: steady state}
\end{figure}

By analyzing the forces acting on the reflector, we can obtain the following equation of motion:

\begin{equation}
\begin{aligned}\label{eq: acceleration}
    & \ddot{x}=\left (  \frac{10 \dot{h}^2}{14 h^{2}} - \frac{2 \ddot{y}}{5h}-\frac{4\nu \dot{h}}{h^{3}} - \frac{C_{e}\dot{h}^2}{4 h^{2}}-Mg \right )\bigg/ (M+\frac{2}{5h})
    \\
   & \text{where } M=\frac{m}{\pi^2 R^4 \rho} \text{  ,  } \nu=\frac{\mu}{\rho}
\end{aligned}\tag{9}
\end{equation}

where the mass ratio ($M$) 
is a measure of the relative significance of the inertial effects of the free floating object and the liquid film, and the kinematic viscosity ($\nu$) is a measure of the fluid resistance to flow under inertial forces. 
 
To understand the nature of this dynamical system, we conducted experiments in which we measured the levitation displacement of a free-floating mass placed on top of a vibrating surface (40~kHz) in a liquid container at different vibration amplitudes, similar to other experiments from literature \cite{Minikes2003,Liu2009,Wang2021}. As demonstrated in Fig.\ref{fig: steady state}(a) for a mass of (64~g), experiments confirmed the existence of stable squeeze-film levitation in liquid 
(refer to the supplemental material \cite{supp} for details about the experimental setup).

By looking closely at the response shown in Fig.\ref{fig: steady state}(b), we observe the following: the reflector reaches an equilibrium levitation position while oscillating. We notice that the amplitude of the reflector oscillations is considerably smaller than the input amplitude ($\leq 15\% $). 
In addition, we also find the phase shift between the reflector oscillations and input vibrations to be consistently around ($\frac{\pi}{2}$). Finally, the dynamic response of the system has two characteristic time scales: (a) time corresponding to the mass reaching a steady-state levitation position $t_{ss}$ (b) time associated with the reflector oscillations $t_{fl}$, such that ($t_{ss} \gg t_{fl}$). 
We can exploit this observation to decompose the response into two components; namely, a time-averaged and an oscillating components as follows: 

\begin{equation}
    x\ =\ \bar{x}+\tilde{x} \ =\ \bar{x}+\gamma \cos(\omega t + \phi)
    \tag{10}
    \label{eq: x average}
\end{equation}
where $\bar{x}$ is the time-averaged steady-state levitation distance and $\tilde{x}$ is the oscillation component such that $\tilde{x}=\gamma \cos(\omega t +\phi)$. 
Since we are mainly interested in the time-averaged levitation component ($\bar{x}$), we can impose a time average operator $\langle . \rangle$ on the dynamic equation Eq.\ref{eq: acceleration}. 
This time average operator is given by $\langle . \rangle = \frac{1}{T} \int_{0}^{T} .\quad dt$, where $T$ is the period of the oscillation. 
The analytical time-average of Eq.\ref{eq: acceleration} exists in the special case when the mass ratio $M$ is negligible compared to the $\frac{2}{5h}$ term in the denominator of the right hand side of Eq.\ref{eq: acceleration}. We can find the range of values where the assumption of a negligible mass ratio is valid to be as follows: for masses of few hundred grams $O(10^{-1})$, radii of tenth of millimeters $O(10^{-2})$ and for a typical liquid density of order $O(10^{3})$, the mass ratio is of order $O(10^{3})$ compared to an $O(10^{5})$ for the $\frac{2}{5h}$ term. 
By imposing this assumption on Eq.\ref{eq: acceleration}, we obtain the following expression:

\begin{equation}
\begin{aligned}\label{eq: assumption}
   \ddot{x}= \frac{25 \dot{h}^2}{14 h} - \ddot{y}-\frac{10\nu \dot{h}}{h^{2}} - \frac{5 C_{e} \dot{h}^2}{8 h}- \frac{5hMg}{2}
\end{aligned}\tag{11}
\end{equation}

\begin{figure}
    \includegraphics{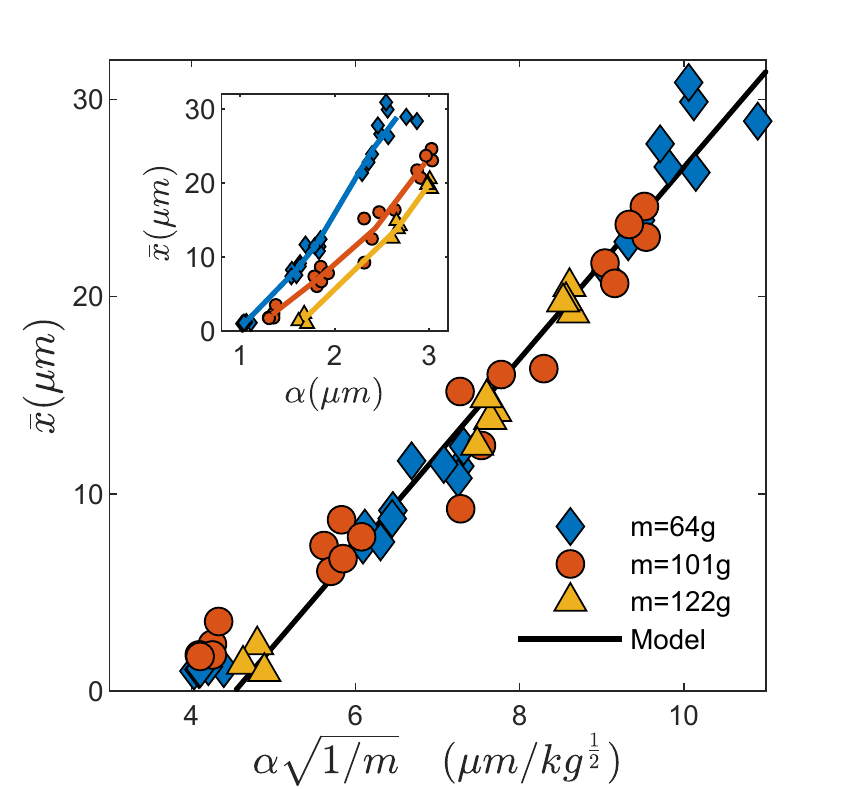}
    \caption{Validation of the model demonstrated by the steady-state time-averaged levitation height $\bar{h}$ as a function of the vibration amplitude and mass of the reflector expressed as $\alpha \sqrt{1/m}$. The experiments were conducted on three different masses: 64~g (blue parallelograms), 101~g (orange circles) and 122~g (yellow triangles) at vibration amplitudes up to $3\mu m$. We identified the initial film thickness $h_{0}$ to be $\approx 22 \mu m$ and the pressure loss coefficient $C_{e}$ to be $\approx 9$. The model (black solid line) shows close agreement with the experimental data (individual plotted points) at different combinations of vibration amplitudes and masses. The inset shows the decomposed experimental data; the steady-state time-averaged levitation height $\bar{h}$ as a function of the vibration amplitude $\alpha$ for each mass individually.}
    \label{fig:model validation}
\end{figure}

A remarkable consequence of the $M \ll \frac{2}{5h}$ assumption is that  all of the terms of Eq.\ref{eq: acceleration} become symmetric in Eq.\ref{eq: assumption} except for the convection terms $\frac{25 \dot{h}^2}{14 h}$ and $\frac{5 C_{e} \dot{h}^2}{8 h}$. It means that within the range of values in which the assumption is valid, the convection effects prevail and becomes the sole source of the steady-state time-averaged levitation. By imposing the time average operator, the symmetric terms $\ddot{x}$, $\ddot{y}$ and $\frac{10 \nu \dot{h}}{h^2}$ of Eq.\ref{eq: assumption}, by definition, converge to zero which yields the following:

\begin{equation}
\begin{aligned}\label{eq: time-average}
   \langle h^2 \rangle = \frac{(20-2.2 C_{e})}{28Mg} \langle \dot{h}^2 \rangle
\end{aligned}\tag{12}
\end{equation}
The time average of the film height yields the steady-state component $\bar{h}$. 
On the other hand, we can find the time average of the $\dot{h}^2$ by substituting the derivative of Eq.\ref{eq: x average} into the $\dot{h}$ expression of Eq.\ref{eq: film thickness} and find the time average integral, to be:

\begin{equation}
\begin{aligned}\label{eq: time-average-hdot}
   \langle \dot{h}^2 \rangle = \frac{\alpha^2 \omega^2}{2} + \frac{\gamma^2 \omega^2}{2} - \alpha \gamma \omega^2 \cos(\phi)
\end{aligned}\tag{13}
\end{equation}

Given our earlier experimental findings that the amplitude of the reflector oscillations $\gamma$ is  ($\leq~15\%$) of the input amplitude and that the phase shift $\phi$ is around ($\frac{\pi}{2}$), we can conclude that the reflector dynamics terms $\gamma^2 \omega^2/2$ and $\alpha \gamma \omega^2 \cos(\phi)$ are negligible compared to the vibration input term $\alpha^2 \omega^2/2$. By omitting the negligible terms from Eq.\ref{eq: time-average-hdot}, substituting it back in Eq.\ref{eq: time-average} and expanding the mass ratio $M$, we find the following expression for the time-averaged steady-state film height $\bar{h}$:

\begin{equation}
\begin{aligned}\label{eq: time-average-final-formula}
    \bar{h} = \pi R^2 \alpha \omega \, \Phi\, \sqrt{\frac{\rho}{mg}} 
\end{aligned}\tag{14}
\end{equation}
where $\Phi(C_{e})$ is a correction factor that accounts for the energy dissipation due to the edge effect, such that $\Phi=\sqrt{(20-2.2 C_{e})/56}$. 
To validate the theory, we conducted experiments using different combinations of masses ($64$, $101$ and $122$ g) with a disk radius (10~mm) at ultrasonic (40~kHz) vibration amplitudes ($1$-$3$ $\mu m$),  in a purified liquid of density (1030~kg/m$^3$). 
Using the collected data, we identified the empirical parameters ($h_0,C_e$) to be $h_0\approx 22\mu m$ and  $C_e\approx 9$. As demonstrated in Fig.\ref{fig:model validation}, the model shows close agreement with the experimental data. We also observe that the correction factor $\Phi$ is not influenced by the vibration amplitude nor the mass, as expected, since the loss coefficient depends only on the geometry of the film entrance. In addition, we see that levitation starts at certain amplitude threshold, similar to air systems \cite{Wang2021}, which 
is mainly because for thicker squeeze films, higher amplitude is required to induce sufficient pressure in the film to outweigh the mass of the reflector and vice versa.


In summary, we showed that liquid inertia, manifested as the temporal and convective acceleration of the liquid film, is the source of pressure asymmetry that causes the levitation phenomenon to exist. We demonstrated that in the special case of a negligible mass ratio ($M$) compared to a film thickness parameter ($\frac{2}{5h}$), the fluid convective inertial effects prevail and become the sole source of the levitation phenomenon. In such a case, we derived a closed form compact formula that predicts the steady-state time-averaged levitation height as a function of the vibration input parameters ($\alpha , \omega$), liquid density ($\rho$) and mass ($m$) of the levitated object. 
The findings of this letter could potentially lead to novel physics-informed designs of in-liquid levitation systems for non-contact manipulation, transportation and friction modulation applications.
\\

The data that support the findings of this study are openly
available in 4TU.ResearchData Repository \cite{Atalla2022}.

\providecommand{\noopsort}[1]{}\providecommand{\singleletter}[1]{#1}%

\end{document}